\documentclass[english,prb,aps,amsmath,amssymb,reprint,superscriptaddress,showkeys]{revtex4-2}
\usepackage{mathptmx}
\usepackage[LGR,T1]{fontenc}
\usepackage[utf8]{inputenc}
\usepackage{color}
\usepackage{babel}
\usepackage{array}
\usepackage{multirow}
\usepackage{amsmath}
\usepackage{amssymb}
\usepackage{graphicx}
\usepackage{wasysym}
\usepackage[bookmarks=true,bookmarksnumbered=false,bookmarksopen=false,
 breaklinks=true,pdfborder={0 0 0},pdfborderstyle={},backref=false,colorlinks=true]
 {hyperref}
\hypersetup{pdftitle={First principles modeling of group-III acceptors and their potential
lifetime-limiting effects in n-type 4H-SiC},
 pdfauthor={José Coutinho},
 allcolors=blue}

\makeatletter

\DeclareRobustCommand{\greektext}{%
  \fontencoding{LGR}\selectfont\def\encodingdefault{LGR}}
\DeclareRobustCommand{\textgreek}[1]{\leavevmode{\greektext #1}}

\providecommand{\tabularnewline}{\\}

\usepackage{orcidlink}

\makeatother

\begin{document}
\selectlanguage{english}
\title{First principles modeling of group-III acceptors and their potential
lifetime-limiting effects in n-type 4H-SiC}
\author{José Coutinho\orcidlink{0000-0003-0280-366X}}
\affiliation{i3N and Department of Physics, University of Aveiro, Campus Santiago,
3810-193 Aveiro, Portugal}
\email{jose.coutinho@ua.pt}

\begin{abstract}
Accurate modeling of p-type dopants in 4H-SiC is essential for understanding
the mechanisms governing doping efficiency and carrier transport.
In this work, we revisit the electronic structure of boron- and aluminum-related
acceptors using hybrid density functional methods. Besides defect
formation energies and thermodynamic transition levels, we present
a quantitative look into the carrier capture kinetics within the multi-phonon
emission framework. Our results reveal striking differences between
the two most relevant p-type dopants. While $\textrm{B}_{\textrm{Si}}$
and $\textrm{B}_{\textrm{C}}$ exhibit nearly identical formation
energies, consistent with the occurrence of both defects, the formation
energy of $\textrm{Al}_{\textrm{C}}$ under intrinsic conditions is
approximately 6.5 eV higher than that of $\textrm{Al}_{\textrm{Si}}$,
confirming previous findings that the former is unlikely to occur.
We further find that $\textrm{B}_{\textrm{Si}}$ possesses large electron
and hole capture cross sections, identifying it as a plausible source
of minority-carrier lifetime degradation of n-type material contaminated
with boron. In addition, we predict a previously unexplored donor
transition for $\textrm{B}_{\textrm{C}}$. If experimentally confirmed,
this defect would represent additional problems to both p-type doped
and boron contaminated 4H-SiC, not only because of its ineffectiveness
as an electric dopant, but also due to trapping of up to two free
holes, reducing the free-hole concentration and increasing scattering
effects.

\vspace{0.5cm}
\noindent \emph{© 2026 Author(s). All article content, except where otherwise noted, is licensed under a Creative Commons Attribution (CC BY) license (\href{https://creativecommons.org/licenses/by/4.0/}{https://creativecommons.org/licenses/by/4.0/}). \href{https://doi.org/10.1063/5.0351340}{https://doi.org/10.1063/5.0351340}}
\end{abstract}
\maketitle

\section{Introduction\label{sec:intro}}

Silicon carbide (SiC) is a frontrunner semiconductor for high-power
electronics. Landmarking achievements occurred in 2001 with the market
launch of the first SiC-based Schottky diode, and in 2011 when the
first SiC metal-oxide field effect transistor (MOSFET) was made available.
For such devices, the hexagonal 4H phase outperforms other polytypes
(3C and 6H), featuring higher and nearly-isotropic carrier mobility,
wider band gap and higher breakdown voltage \citep{Kimoto2014,Langpoklakpam2022,Kimoto2022}.

When compared to its rival GaN \citep{Narita2020}, SiC offers simpler
processing in order to realize both n-type and p-type substrates and
implants. However, the performance of p-dopants (group-III acceptors)
falls behind that of their n-type counterparts (nitrogen and phosphorus).
Nitrogen has an ionization energy of about 60 or 125~meV (depending
on the sub-lattice location of N, \emph{i.e.} cubic or hexagonal)
\citep{Ikeda1980,Ivanov2005}, leading to complete or nearly complete
ionization at room temperature, even for a nitrogen concentration
as high as $[\textrm{N}]\lesssim10^{17}\,\textrm{cm}^{-3}$ \citep{Kimoto2014}.

Owed to their relatively high hole ionization energies ($\sim\!0.2\textrm{-}0.3\:\textrm{eV}$),
group-III acceptors suffer from severe incomplete ionization (see
Appendix~A of Ref.~\citep{Kimoto2014}). The use of boron is usually
avoided due to formation of an additional and deeper boron-related
hole trap (referred to as D-center at $E_{\textrm{v}}+0.7\:\textrm{eV}$)
\citep{Anikin1985,Suttrop1990,Troffer1997}, as well as for its anomalously
high out-diffusivity \citep{Negoro2007}. Aluminum on the other hand,
outperforms the other elements, but still shows a room-temperature
ionization fraction of only 30\% for $[\textrm{Al}]=10^{16}\:\textrm{cm}^{-3}$,
which drops below 10\% for $[\textrm{Al}]\gtrsim10^{17}\,\textrm{cm}^{-3}$
\citep{Darmody2019}. Additional problems include incomplete activation
and compensation effects of dopant implants (even after subjecting
the material to high-temperature thermal treatments), difficulties
in the fabrication of Ohmic contacts, as well as limited hole mobility
due to scattering and trapping effects by the relatively deep group-III
acceptor levels \citep{Handy2000,Saks2001,Heera2001,Kimoto2014,Darmody2019}.
These are some of the hurdles that have hampered the fabrication of
p-type and bipolar SiC devices.

Electrically active defects are often divided into \emph{trapping}
and \emph{recombination} centers. Trapping/detrapping of carriers
leads to quasi-static charge accumulation, diverts electric field
lines, and blockades drain current. In practice this limits the high-frequency
power output of the SiC devices \citep{Binari2002}. Recombination
centers on the other hand, interact quickly with both types of carriers,
leading to a degradation of minority carrier lifetime (hereafter referred
to simply as \emph{lifetime}), with important repercussions to the
on-state resistance of bipolar devices \citep{Kimoto2022}, or to
the efficiency of SiC radiation detectors \citep{Coutinho2021b,Ruddy2022}.

In n-type 4H-SiC, the lifetime is usually tied to the concentration
of a prominent electron trap at $E_{\textrm{c}}-0.7\:\textrm{eV}$,
known as $\textrm{Z}_{1/2}$ and ascribed to carbon vacancy ($\textrm{V}_{\textrm{C}}$)
\citep{Hemmingsson1998,Klein2006,Danno2007,Yamashita2020,Erlekampf2021}.
The concentration of $\textrm{V}_{\textrm{C}}$ defects can be largely
suppressed ($[\textrm{V}_{\textrm{C}}]\apprle10^{11}\:\textrm{cm}^{-3}$)
upon injection/annealing of C interstitials into SiC epi-layers (\emph{e.g.}
carbon implantation/annealing or thermal oxidation \citep{Storasta2007,Hiyoshi2009,Ayedh2015}).
Lifetimes as high as 30~\textgreek{μ}s have been reported after such
treatments \citep{Ichikawa2012}.

As $\textrm{V}_{\textrm{C}}$ becomes less relevant, other traps become
relatively more harmful \citep{Erlekampf2021}. Boron centers have
been connected to lifetime limitations in n-type 4H-SiC \citep{Storasta2002,Ghezellou2023}.
The work of Ghezellou \emph{et~al.} \citep{Ghezellou2023} shows
that when the concentration of $\textrm{Z}_{1/2}$ drops below $[\textrm{Z}_{1/2}]\sim10^{11}\textrm{-}10^{12}\:\textrm{cm}^{-3}$,
an inverse correlation is found between the lifetime and the concentration
of the D-center, and not with $[\textrm{Z}_{1/2}]$. It also shows
that the detrimental effect of the D-center at room temperature, is
mitigated by the trapping/detrapping kinetics involving the shallower
boron acceptor.

Boron contamination of SiC layers seems difficult to avoid. It possibly
originates from graphite components of growth reactors \citep{Storasta2002,Ghezellou2023}.
It is therefore highly desirable to understand the kinetics of carrier
trapping at group-III impurities, which could be unavoidable contaminants
in n-type material and limit the lifetime, but also in p-type material,
where degradation effects pertain to carrier scattering and charging
effects. It is also important to estimate these properties at device
operating conditions, and not only at those temperatures allowed by
the experimental conditions. For that, recent advances in first-principles
calculations of carrier capture rates, stand as a useful tool \citep{Shi2012,Alkauskas2014,Kim2019}.
However, and before heading that way, we leave some introductory notes
regarding the meaning of quantities measured by junction spectroscopy,
and how they relate to the calculations.

The rate at which free carriers are captured by defect traps, for
example hole traps \citep{Peaker2018}, is usually written as

\noindent
\begin{equation}
c=\sigma\,\langle v_{\textrm{v}}\rangle\,p=C\,p.\label{eq:c_ht}
\end{equation}
The expression in the middle translates the exposure of an empty hole
trap with \emph{capture cross section} $\sigma$ (a concept borrowed
from nuclear physics), to a steady-state flux $\langle v_{\textrm{v}}\rangle p$
of holes with density $p$, traveling with average thermal velocity
$\langle v_{\textrm{v}}\rangle$. The subscript `v' indicates that
we are dealing with a property that relates to the valence band top.
The expression on the right makes use of an alternative concept, namely
the capture coefficient $C=\sigma\,\langle v_{\textrm{v}}\rangle$
(\emph{i.e.} the capture rate per unit of carrier density). As explained
in the seminal paper of Alkauskas and co-workers \citep{Alkauskas2014},
this form is preferred from a theoretical perspective since $C$ can
be directly calculated (without considering the carrier velocity).
Analogous quantities can be defined for the capture rate of electrons
traveling in the conduction band bottom with average thermal velocity
$\langle v_{\textrm{c}}\rangle$, and electron density $n$.

\noindent In thermal equilibrium, the capture and emission of carriers
must satisfy the principle of detailed balance, so that hole traps
must also return holes back to the valence band top at an emission
rate

\noindent
\begin{equation}
e=\sigma\langle v_{\textrm{v}}\rangle N_{\textrm{v}}\exp\left(-\frac{\Delta G}{k_{\textrm{B}}T}\right),\label{eq:e_ht}
\end{equation}
where $N_{\textrm{v}}$ is the effective density of states at the
valence band top, and $\Delta G$ is the Gibbs free energy of ionization
of the defect upon hole emission.

The meaning of $\sigma$ and $\Delta G$ have been discussed at length
elsewhere \citep{Stoneham1981,Blood1992,Shi2012,Alkauskas2014}. We
highlight the following difficulties in the analysis of capture/emission
processes by defect traps: (1) all quantities in Eqs.~\ref{eq:c_ht}
and \ref{eq:e_ht} (except the Boltzamann constant, $k_{\textrm{B}}$)
are temperature-dependent. Notably, a cross section measured at $T=100\:\textrm{K}$
can differ by several decades from the same quantity at room temperature;
(2) $\Delta G$ is a thermodynamic property which defines a transition
energy level above the valence band top, measured as a hole trap (for
an electron trap it would refer to a level below the conduction band
bottom). Its location with respect to the Fermi level (at a certain
temperature) defines the equilibrium populations of filled/empty traps;
(3) $\sigma$ is a kinetic quantity that embodies several rate-limiting
properties, including transition matrix elements between free to bound
states, electron phonon coupling, or the effect of a capture barrier.
(4) In the high temperature limit, the capture cross section follows
a classical Arrhenius law, $\sigma(T)=\sigma_{\infty}\exp(-E_{\sigma}/k_{\textrm{B}}T)$,
where $E_{\sigma}$ is the capture barrier, and $\sigma_{\infty}$
is the cross section extrapolated to $T\rightarrow\infty$. However,
at low temperatures, the carrier capture can be dominated by quantum
mechanical tunneling, and if used under such conditions, the above
Arrhenius relation severely underestimates the cross section.

The standard DLTS procedure involves sending a filling pulse through
a diode to populate defect traps with charge carriers, followed by
application of reverse-bias to create a depletion region. This will
place the diode in a non-equilibrium state, allowing for monitoring
the time-evolution of trap emptying (via measurement of the capacitance,
current, etc.), which is governed by a transient proportional to $1-\exp(-e\,t)$
\citep{Henry1977,Blood1992,Peaker2018}. It is assumed that $\langle v_{\textrm{v}}\rangle\,N_{\textrm{v}}=\gamma_{\textrm{v}}\,T^{2}$
($\gamma_{\textrm{v}}$ being a material-related constant \citep{Blood1992}).
Then, one plots $\ln\left(e/T^{2}\right)$ against $1/T$, and fit
the following Arrhenius relation,

\noindent
\begin{equation}
e/T^{2}=\gamma_{\textrm{v}}\,\sigma_{\textrm{a}}\,\exp\left(-\frac{E_{\textrm{a}}}{k_{\textrm{B}}T}\right),\label{eq:e_ht-arrhenius}
\end{equation}
to the data in order to extract the \emph{apparent} cross section
($\sigma_{\textrm{a}}$) and activation energy for emission ($E_{\textrm{a}}$).
Importantly, like $\sigma$ and $\Delta G$ in Eq.~\ref{eq:e_ht},
the $\sigma_{\textrm{a}}$ and $E_{\textrm{a}}$ parameters in Eq.~\ref{eq:e_ht-arrhenius}
are also temperature-dependent. Despite that, the fit works well because
the measured data is usually limited to a narrow temperature range
where the semi-log plots are apparently linear \citep{Wickramaratne2018}.
On the other hand, this means that (i) $\sigma_{\textrm{a}}$ can
be affected by large errors, (ii) it cannot be decomposed into clear
physical quantities, and (iii) it bears little resemblance either
to $\sigma$ or to the classical limit $\sigma_{\infty}$. A direct
measurement of $\sigma(T)$ is however possible by following the trap
filling kinetics upon injection. Unfortunately, the temperature range
of such measurements is also limited, depending on several factors,
such the thermal stability of the defect, or the heigh of the capture
barrier \citep{Peaker2018}.

Difficulties also exist in the interpretation of the activation energy.
For instance, given that $\Delta G=\Delta H-T\Delta S$, and that
at low temperatures the ionization enthalpy ($\Delta H$) is weakly
dependent on $T$, the extracted activation energy for emission becomes
$E_{\textrm{a}}\approx\Delta H$. However, at higher temperatures,
and in the presence of a capture barrier, the emission/capture rates
are now affected by a factor of $\exp(-E_{\sigma}/k_{\textrm{B}}T)$,
and the extracted activation energy becomes $E_{\textrm{a}}\approx\Delta H+E_{\sigma}$.
The above was recently pointed out in Ref.~\citep{Wickramaratne2018},
emphasizing that great care should be exercised when comparing measured
electrical levels and capture cross sections with calculated analogues.

Finally, in the present context, we also note that electronic structure
calculations, mostly based on density functional theory, usually neglect
finite temperature effects, and the transition levels are simply given
by $\Delta G(T=0\:\textrm{K})$. In that respect, the calculated transition
levels are better compared to the emission enthalpy at $T=0\:\textrm{K}$.
Of course, a more elaborate approach would include vibrational and
electronic contributions to the free energy difference $\Delta G$,
which would involve the calculation of these effects for both states
(before and after carrier emission) \citep{Wickramaratne2018}.

In this paper we use modeling tools based on first-principles electronic
structure methods, to investigate the interaction between charge carriers
and group-III dopants in 4H-SiC. We confirm that the most probable
boron and aluminum acceptor impurities in 4H-SiC are very different
in terms of defect geometries and site locations, but we also unravel
staggering differences regarding their carrier capture/emission kinetics.
The work is based on hybrid density functional calculations of the
electronic signature of several Al and B centers, including their
transition levels and capture coefficients/cross-sections. Another
major finding is that the B impurity replacing a carbon atom has a
donor transition (besides the already known acceptor). This result
should prompt us to reconsider the effects of B contamination to hole
trapping/detrapping and lifetime limitations of 4H-SiC.

The next sections of the paper are organized in the following way:
After detailing the methods employed in Sec.~\ref{sec:methods},
we report the results in Sec.~\ref{sec:results}, which is divided
in two subsections. The first deals with the stability of B and Al
defects, including the calculation of electronic transitions. In the
second we address the calculation of the kinetic properties of the
transitions. We end with the conclusions in Sec.~\ref{sec:conclusions}.

\section{Methods\label{sec:methods}}

First-principles calculations were carried out using the density functional
Vienna \emph{Ab initio} Simulation Package (VASP) \citep{Kresse1993,Kresse1994,Kresse1996,Kresse1996a},
employing the projector-augmented wave method for the treatment of
core states \citep{Blochl1994}. A basis set of plane-waves with kinetic
energy of up to 400~eV was used to describe the Kohn-Sham states.
The total energy, electron density, electronic states and band structure
of defective supercells were evaluated self-consistently, using the
screened hybrid density functional of Heyd-Scuseria-Ernzerhof (HSE06)
\citep{Heyd2003,Heyd2006}, with a numerical accuracy for the total
energy of $10^{-7}$ eV.

Defects were investigated in 400-atom (defect-free) supercells of
4H-SiC. These were obtained by replication of $5\!\times\!5\!\times\!2$
primitive cells. The equilibrium (calculated) lattice parameters of
the 4H-SiC crystal were $a=3.071$~Å and $c=10.052$~Å, corresponding
to a supercell volume of $V=4105~\textrm{Å}^{3}$. Defect geometries
were consistently optimized based on HSE06-level forces, using a conjugate-gradient
algorithm. The maximum force acting on the nuclei of fully relaxed
defect geometries was lower than 5~meV/Å. All calculations employed
$\Gamma$-sampling of the Brillouin zone (BZ). Additional details
and convergence tests regarding defect energetics and geometry can
be found in Ref.~\citep{Torres2022} and respective supplemental
information. In that work, $\Gamma$-sampled formation energies and
electronic transition levels were compared to analogous $2\!\times\!2\!\times\!2$-sampled
quantities. We found that both results were consistent and differed
by less than 50~meV.

Formation energies ($E_{\textrm{f}}$) and transition levels of defects
(for instance with respect to the valence band top, $\Delta E(q/q')=E(q/q')-E_{\textrm{v}}$)
were respectively evaluated using the usual methodologies \citep{Freysoldt2014},

\begin{equation}
E_{\textrm{f}}(q,E_{\textrm{F}})=E_{\textrm{def}}(q)-\mu_{\textrm{ref}}+q(\epsilon_{\textrm{v}}+E_{\textrm{F}})\label{eq:e_form}
\end{equation}
and

\begin{equation}
E(q/q')-E_{\textrm{v}}=[E_{\textrm{def}}(q)-E_{\textrm{def}}(q')]/(q'-q)-\epsilon_{\textrm{v}}.\label{eq:levels}
\end{equation}

In the above, $E_{\textrm{def}}(q)$ is the total energy of the defective
supercell in charge state $q$ (which already incorporates a periodic
charge correction \citep{Freysoldt2009,Kumagai2014}), and

\begin{equation}
\mu_{\textrm{ref}}=n_{\textrm{Si}}\mu_{\textrm{Si}}+n_{\textrm{C}}\mu_{\textrm{C}}+\mu_{X},\label{eq:mu_ref}
\end{equation}
is a chemical potential of all elements that are present in the supercell,
i.e., $n_{\textrm{Si}}$ silicon atoms, $n_{\textrm{C}}$ carbon atoms,
and one impurity of $X$ species ($X=\{\textrm{B},\:\textrm{Al}\}$).
Elemental chemical potentials were evaluated as $\mu_{\textrm{Si}}=\mu_{\textrm{Si}}^{0}+(\Delta H_{\textrm{f}}^{\textrm{SiC}}-\Delta\mu)/2$,
$\mu_{\textrm{C}}=\mu_{\textrm{C}}^{0}+(\Delta H_{\textrm{f}}^{\textrm{SiC}}+\Delta\mu)/2$,
and $\mu_{X}=\mu_{X}^{0}$, where $\mu_{\textrm{Si}}^{0}$ and $\mu_{\textrm{C}}^{0}$
are atomic energies taken from crystalline silicon and diamond, respectively,
while $\mu_{X}^{0}$ is either the energy per boron or per aluminum
atom in rhombohedral $\alpha$-boron or fcc-aluminum, respectively
(depending if we are dealing with B- or Al-related defects). The heat
of formation of 4H-SiC was evaluated as $\Delta H_{\textrm{f}}^{\textrm{SiC}}=0.62\:\textrm{eV}$,
and that defines the variability for the Si and C chemical potentials,
$|\Delta\mu|\leq\Delta H_{\textrm{f}}^{\textrm{SiC}}$. The upper
and lower limits of $\Delta\mu$ correspond to Si-rich and C-rich
grown crystals.

The term $\epsilon_{\textrm{v}}+E_{\textrm{F}}$ is the average energy
per carrier in the Fermi level, $E_{\textrm{F}}$. It provides a reference
for evaluating the cost of trapping an amount of charge $q$ at the
defect. $E_{\textrm{F}}$ is an independent variable, such that $\epsilon_{\textrm{v}}\leq\epsilon_{\textrm{v}}+E_{\textrm{F}}\leq\epsilon_{\textrm{c}}$,
where $\epsilon_{\textrm{v}}=7.65\:\textrm{eV}$ and $\epsilon_{\textrm{c}}=10.91\:\textrm{eV}$
are respectively the energy of the highest occupied and lowest unoccupied
states of bulk 4H-SiC, which lead to a calculated band gap width of
$E_{\textrm{g}}=\epsilon_{\textrm{c}}-\epsilon_{\textrm{v}}=3.26\:\textrm{eV}$.

Carrier capture kinetics was investigated assuming that trapping at
defects occurs non-radiatively via multi-phonon emission (MPE). This
is the most frequent trap-assisted relaxation mechanism involving
deep states, especially in indirect-band-gap semiconductors such as
4H-SiC \citep{Henry1977,Stoneham1981}. Radiative and other non-radiative
mechanisms, including trap Auger and cascade, are also possible but
they were left outside the scope of this work. Carrier capture coefficients
were evaluated using the formalism developed by Alkauskas \emph{et~al.}
\citep{Alkauskas2014}, as implemented in the \texttt{nonrad} routines
\citep{Turiansky2021},

\begin{equation}
C=\frac{2\pi}{\hbar}f_{\textrm{s}}gVW_{if}^{2}\sum_{m}w_{m}G_{m}(\Delta E),\label{eq:c}
\end{equation}
where $G_{m}(E)$ is the vibrational lineshape function \citep{Stoneham1981},

\begin{equation}
G_{m}(E)=\sum_{n}\left|\left\langle \chi_{im}\left|Q-Q_{0}\right|\chi_{fn}\right\rangle \right|^{2}\delta\left(E+m\hbar\Omega_{i}-n\hbar\Omega_{f}\right).\label{eq:lineshape}
\end{equation}

Equation~\ref{eq:c} derives from Fermi's golden rule within the
single phonon and static approximations, and encodes a transition
between a free carrier away from an empty defect trap (initial electronic
state, $i$), and the same carrier trapped at the defect (final electronic
state, $f$). The energy drop $\Delta E$ is converted into lattice
vibrations. The factor $g$ accounts for any degenerate multiplicity
of landing states, and $V$ is the supercell volume. The summation
in Eq.~\ref{eq:lineshape} runs over vibrational quantum states $\left\{ \chi_{im},\chi_{fn}\right\} $
of the \emph{accepting mode} $Q$, with harmonic mode frequencies
$\left\{ \Omega_{i},\Omega_{f}\right\} $ for the initial and final
electronic states, respectively. The accepting mode is a mass-weighted
collective motion of the nuclei along a vector coordinate between
initial and final state geometries ($Q_{i}-Q_{f}=\Delta Q$), corresponding
to the potential energy drop $\Delta E=E_{i}(Q_{i})-E_{f}(Q_{f})$
\citep{Alkauskas2012}. The phonon matrix elements $\left\langle \chi_{im}\left|Q-Q_{0}\right|\chi_{fn}\right\rangle $
were evaluated by direct numerical integration, around a chosen reference
geometry $Q_{0}$ (either $Q_{0}=Q_{i}$ or $Q_{0}=Q_{f}$), for which
the trapping state was clearly identifiable within the gap of the
band structure.

The temperature dependence of Eq.~\ref{eq:c} comes from the thermal
occupation factor, $w_{m}$, where the index $m$ refers to the vibrational
level of the initial state. This allows for thermal activation over
the capture barrier defined by the crossing between $i$ and $f$
states. The delta function in Eq.~\ref{eq:c} filters energy-conserving
transitions. In practice cubic spline interpolation is employed to
produce a smooth lineshape function to replace $G_{m}(E)$ \citep{Turiansky2021}.

The electron-phonon coupling matrix element, $W_{if}$, was evaluated
from the slope of the electronic matrix element $\langle\psi_{i}(Q_{0})|\psi_{f}(Q)\rangle$.
This quantity was calculated for several geometries along $Q$ around
$Q_{0}$. The scaling factor $f_{\textrm{s}}(T)$ accounts two effects,
namely (1) the enhancement/frustration of $C$ due to Coulomb scattering
of the traveling carrier by the trap (if the latter is charged in
the initial state), and (2) the enhancement/frustration of $W_{if}$
due to the artificial Coulomb perturbation of the free-carrier state
by the defect in the finite supercell (if the latter is evaluated
in a non-neutral charge state). The first effect is accounted for
by the Sommerfeld parameter. The second effect is accounted for by
the ratio of the integrated radial distribution of the band-edge charge
density to that of an homogeneous electronic state (see Ref.~\citep{Turiansky2021}
for details).

With that, the capture cross section for carriers traveling with average
thermal velocity $\langle v\rangle=(3k_{\textrm{B}}T/m^{*})^{1/2}$
was found from $\sigma=C/\langle v\rangle$. Electron and hole effective
masses, $m_{\textrm{e}}^{*}=0.39$ and $m_{\textrm{h}}^{*}=1.02$,
as well as the static dielectric constant $\epsilon_{0}=9.95$ (for
the evaluation of the scaling factor) were taken from experimental
data (see Chapter~2 of Ref.~\citep{Kimoto2014}), and converted
to isotropic quantities by averaging. Calculated analogues could have
been obtained and used without meaningful differences in the results.
In 4H-SiC, hole effective masses show considerable anisotropy. The
estimated impact of either using axial or basal effective masses ($m_{\textrm{h},\perp}^{*}=1.75$
or $m_{\textrm{h},\parallel}^{*}=0.66$) on the calculated capture
cross sections is up to a factor of 2 in the range $T=100\textrm{-}700\:\textrm{K}$.

\section{Results\label{sec:results}}

According to previous theoretical studies of B- and Al-related defects
in SiC \citep{Fukumoto1996,Deak2001,Matsushima2019}, the lowest energy
structures of these centers were found to be either simple substitutional
$X$ acceptors, or $V_{\textrm{C}}X_{\textrm{Si}}$ complexes. The
work of Ref.~\citep{Matsushima2019} is particularly comprehensive,
and although its was carried out for the cubic phase and within the
generalized gradient approximation (GGA), formation energies should
not be far from those obtained for the 4H phase within HSE06 if we
restrict the Fermi level to the lower half of the band gap. This is
justified by two arguments: (i) the valence states involving the bonding
of point defects in both phases are similar, and (ii) the valence
band maximum of 3C-SiC is only 60~meV below that of the 4H phase
\citep{Park2006} --- most of the band gap mismatch ($\sim0.9\:\textrm{eV}$)
is found between conduction band edges --- thus limiting eventual
effects of the GGA underestimated band gap width on defect levels
close to the valence band top. With the above in mind, we focus our
study on $X_{\textrm{Si}}$ and $X_{\textrm{C}}$ substitutions, as
well as $V_{\textrm{C}}X_{\textrm{Si}}$ complexes in 4H-SiC.

\begin{table}
\caption{\label{tab1}Calculated transition level energies of B-related and
Al-related defects in 4H-SiC. All values are given in eV and with
respect to the valence band top ($E_{\textrm{v}}$). Sub-lattice sites
($h$ or $k$) where the impurity is sitting, are indicated within
parentheses. Vacancy-acceptor complexes are labeled with two sub-lattice
indexes, pertaining to the vacant and impurity sites, respectively.}

\begin{ruledtabular}
\begin{tabular}{lcc}
Defect & $E(0/+)-E_{\textrm{v}}$ & $E(-/0)-E_{\textrm{v}}$\tabularnewline
\hline 
$\textrm{B}_{\textrm{Si}}(k)$ &  & 0.38\tabularnewline
$\textrm{B}_{\textrm{Si}}(h)$ &  & 0.32\tabularnewline
$\textrm{Al}_{\textrm{Si}}(k)$ &  & 0.19\tabularnewline
$\textrm{Al}_{\textrm{Si}}(h)$ &  & 0.23\tabularnewline
$\textrm{B}_{\textrm{C}}(k)$ & 0.20 & 0.69\tabularnewline
$\textrm{B}_{\textrm{C}}(h)$ & 0.22 & 0.74\tabularnewline
$V_{\textrm{C}}\textrm{B}_{\textrm{Si}}(kk,hh,kh,hk)$ & 2.31, 2.32, 2.11, 2.21 & 2.88, 2.89, 2.69, 2.79\tabularnewline
$V_{\textrm{C}}\textrm{Al}_{\textrm{Si}}(kk,hh,kh,hk)$ & 2.15, 2.13, 1.97, 2.02 & 2.53, 2.58, 2.48, 2.53\tabularnewline
\end{tabular}
\end{ruledtabular}

\end{table}

\subsection{Stability of B and Al defects\label{subsec:stability}}

\begin{figure}
\includegraphics[width=8.5cm]{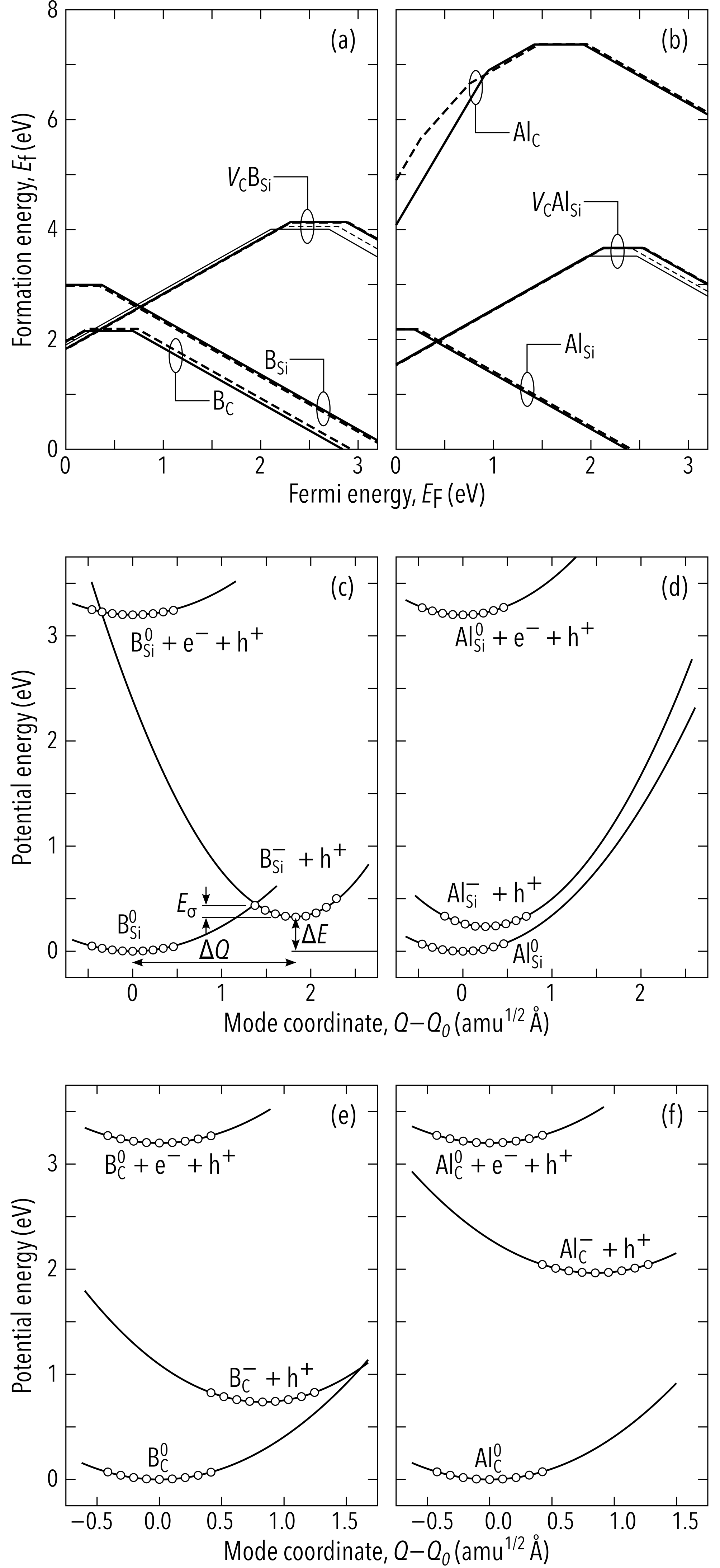}

\caption{\label{fig1}(a, b) Formation energy diagrams for B- and Al-related
defects in 4H-SiC. Results for both substitutional group-III species
at $k$ and $h$ sites are represented by thick solid and dashed lines,
respectively. For $V_{\textrm{C}}X_{\textrm{Si}}$ defects ($X=\{\textrm{B, Al}\}$),
$kk$ and $hh$ axial configurations are represented by thick solid
and dashed lines, while $kh$ and $hk$ alignments are shown as thin
solid and dashed lines, respectively. The calculations assume Si-rich
conditions. (c-f) Configuration coordinate diagrams for carrier capture
by $X_{\textrm{Si}}$, $X_{\textrm{C}}$ defects at $h$-sites. The
meaning of the mode coordinate displacement, $\Delta Q$, capture
barrier, $E_{\sigma}$, and trap depth, $\Delta E$, upon hole capture
are depicted in (c) for the case of $\textrm{B}_{\textrm{Si}}^{-}(h)+h^{+}\rightarrow\textrm{B}_{\textrm{Si}}^{0}(h)$.}
\end{figure}

We start by analyzing the relative stability of the defects of interest
as a function of the Fermi level. Figures~\ref{fig1}(a) and \ref{fig1}(b)
depict formation energy diagrams for B- and Al-related defects in
4H-SiC, respectively. For the simple substitutional impurities we
investigated their properties on both $k$ (solid lines) or $h$ (dashed
lines) sub-lattice sites of the crystal. For $V_{\textrm{C}}X_{\textrm{Si}}$
complexes, we report results for all four alignments, namely axial
$kk$ and $hh$, as well as basal $kh$ and $hk$, where the first
and second sub-lattice indexes pertain to the vacant and impurity
sites, respectively.

The results shown in Figs.~\ref{fig1}(a) and \ref{fig1}(b) were
obtained for silicon-rich conditions. For silicon-poor conditions,
$E_{\textrm{f}}$ values of $X_{\textrm{C}}$/$X_{\textrm{Si}}$ defects
are shifted upward/downward by $\Delta E_{\textrm{SiC}}^{\textrm{f}}=0.62\:\textrm{eV}$.
Formation of $V_{\textrm{C}}X_{\textrm{Si}}$ complexes involve the
subtraction of a Si-C pair of atoms. That leaves the Si/C concentration
ratio in the supercell unchanged, and therefore their formation energy
is insensitive to the inter-dependence between Si and C chemical potentials.

In line with previous reports \citep{Fukumoto1996,Petrenko1996,Deak2001,Matsushima2019,Torres2022},
we find that both $\textrm{B}_{\textrm{Si}}$ and $\textrm{B}_{\textrm{C}}$
impurities show rather low formation energies, and therefore are likely
to form during growth. Figure~\ref{fig1}(a) shows how close the
formation energies of $\textrm{B}_{\textrm{Si}}$ and $\textrm{B}_{\textrm{C}}$
are, particularly in their negative states. These are expected to
be most favored under intrinsic and growth conditions, when the Fermi
level is close to mid-gap. While $\textrm{B}_{\textrm{C}}$ is more
stable under Si-rich conditions (as depicted in Fig~\ref{fig1}(a)),
for Si-poor grown material $\textrm{B}_{\textrm{Si}}$ is naturally
favored, becoming more stable than $\textrm{B}_{\textrm{C}}$. This
issue has been discussed at length elsewhere \citep{Fukumoto1996,Petrenko1996,Deak2001,Matsushima2019,Torres2022}.

The relative stability of $V_{\textrm{C}}\textrm{B}_{\textrm{Si}}$
with respect to substitutional impurities is also interesting. The
relevance of this issue stems from an interchangeable assignment of
$\textrm{B}_{\textrm{C}}$ and $V_{\textrm{C}}\textrm{B}_{\textrm{Si}}$
to a deep DLTS boron-related trap at about $E_{\textrm{v}}+0.6\:\textrm{eV}$,
and referred to as D-center \citep{Baranov1998,DuijnArnold1998}.
In a recent paper \citep{Torres2022}, we reported that besides the
agreement between the calculated acceptor level of $\textrm{B}_{\textrm{C}}$
and the trap depth of the D-center, only the calculated $g$-tensor
and hyperfine coupling constants of $\textrm{B}_{\textrm{C}}$ were
compatible with the available magnetic resonance experimental data
connected to the same center \citep{Baranov1998,DuijnArnold1998}.

Figure~\ref{fig1}(a) shows that $V_{\textrm{C}}\textrm{B}_{\textrm{Si}}$
has donor and acceptor transitions in the upper half of the gap. The
list of transitions for all possible orientations are provided in
Tab.~\ref{tab1}. The results are consistent with previous theoretical
findings \citep{Deak2001}, and confirm that (i) $V_{\textrm{C}}\textrm{B}_{\textrm{Si}}$
cannot account for a transition level in the lower half of the gap
as expected for the D-center; (ii) $V_{\textrm{C}}\textrm{B}_{\textrm{Si}}$
complexes, having a relatively high formation energy, are not so likely
to form during growth (intrinsic conditions) in comparison to $\textrm{B}_{\textrm{Si}}$
and $\textrm{B}_{\textrm{C}}$; (iii) $V_{\textrm{C}}\textrm{B}_{\textrm{Si}}$
(and $V_{\textrm{C}}\textrm{Al}_{\textrm{Si}}$) could act as compensation
centers in p-type material, effectively subtracting two holes per
complex formed --- one due to deactivation of the group-III impurity,
plus one due to capture by the deep donor state. We did not find additional
donor transition levels for $V_{\textrm{C}}\textrm{B}_{\textrm{Si}}$
(nor $V_{\textrm{C}}\textrm{Al}_{\textrm{Si}}$).

Boron on the Si site is confirmed as being an acceptor impurity, consistent
with the \emph{shallow boron} DLTS peak at $E_{\textrm{v}}+0.27\:\textrm{eV}$
\citep{Sridhara1998,Zhang2003}. The calculated acceptor transitions
for $\textrm{B}_{\textrm{Si}}(k)$ and $\textrm{B}_{\textrm{Si}}(h)$
are found at $E_{\textrm{v}}+0.38\:\textrm{eV}$ and $E_{\textrm{v}}+0.32\:\textrm{eV}$,
respectively (see Table~\ref{tab1}). Details of defect geometries
were reported in Refs.~\citep{Gerstmann2004,Torres2022}. Here we
leave the main findings: The geometry of the neutral states ($\textrm{B}_{\textrm{Si}}^{0}(k)$
and $\textrm{B}_{\textrm{Si}}^{0}(h)$) consists of an off-site B
atom with three-fold coordination, giving rise to a localized radical
whose wavefunction is essentially a p-orbital on a C atom along one
of the basal B-C directions. This is the state were the hole becomes
trapped. However, in the negative state (upon hole emission), a strong
reconfiguration takes place and the B atom becomes four-fold coordinated.
Interestingly, the calculated band structure of the negative states
($\textrm{B}_{\textrm{Si}}^{-}(k)$ and $\textrm{B}_{\textrm{Si}}^{-}(h)$)
shows a clean band gap, thus indicating that the acceptor transition
of $\textrm{B}_{\textrm{Si}}$ has a polaronic nature, \emph{i.e.}
the capture of a hole by $\textrm{B}_{\textrm{Si}}^{-}$ must be accompanied
by a strong local distortion.

The calculated acceptor levels for $\textrm{B}_{\textrm{C}}(k)$ and
$\textrm{B}_{\textrm{C}}(h)$ are found at $E_{\textrm{v}}+0.69\:\textrm{eV}$
and $E_{\textrm{v}}+0.74\:\textrm{eV}$, respectively. Here the B
atom is approximately located on crystallographic C-sites. These values
are consistent with the deep boron-related DLTS peak corresponding
to a level at about $E_{\textrm{v}}+0.6\:\textrm{eV}$ \citep{Anikin1985,Suttrop1990}.
$\textrm{B}_{\textrm{C}}$ defects create a doubly-degenerate state
deep in the gap, which is partially occupied in the neutral state,
$e^{\uparrow\uparrow\downarrow+}$, where the superscripted $+$ sign
represents a trapped hole. That configuration leads to a minute Jahn-Teller
displacement of the B atom ($<0.1$~Å) along the $\langle0\bar{1}10\rangle$
basal direction. In the negatively charge state, the electronic doublet
becomes fully occupied and $\textrm{B}_{\textrm{C}}^{-}$ recovers
the perfect trigonal symmetry.

The depth and localization of the $e^{\uparrow\uparrow\downarrow+}$
gap states of $\textrm{B}_{\textrm{C}}^{0}$ led us to hypothesize
the existence of donor activity. Indeed we find donor transitions
for $\textrm{B}_{\textrm{C}}(k)$ and $\textrm{B}_{\textrm{C}}(h)$
at $E_{\textrm{v}}+0.20\:\textrm{eV}$ and $E_{\textrm{v}}+0.22\:\textrm{eV}$,
respectively. We note that the calculated donor states, $\textrm{B}_{\textrm{C}}^{+}(k)$
and $\textrm{B}_{\textrm{C}}^{+}(h)$, show trigonal symmetry owed
to a localized $e^{\uparrow\uparrow++}$ doublet within the gap (total
spin $S=1$). Considering the depth of the calculated levels along
with the usual error bar of the method ($\sim0.1$~eV), we must consider
this finding reliable.

However, bear in mind that the error bar could widened, especially
if vibrational entropy changes are important (not to mention configurational
and electronic free energy changes upon ionization, which are in principle
are negligible). Additionally, as detailed in Sec.~\ref{sec:methods},
there is a temperature-dependence of the capture coefficient which
effectively ``bends'' the DLTS data with respect to the ideal Arrhenius
plot. That affects the extracted value of the emission energy, considering
that experiments are done within a limited temperature range. It has
been shown in Ref.~\citep{Wickramaratne2018} that both effects could
shift the calculated transition levels with respect to those evaluated
at $T=0\:\textrm{K}$ by up to about 0.2~eV. Hence, great care and
prudence should be employed when performing direct comparisons between
calculated transition levels and activation energies extracted from
DLTS.

The unintentional presence of boron in 4H-SiC and the eventual formation
of $\textrm{B}_{\textrm{C}}$, has been considered a potential source
of problems for both p-type and n-type materials. In n-type 4H-SiC
such concerns are rooted on the fact that the center is negatively
charged, that it could trap minority carriers (up to two holes according
to the above finding), thus possibly limiting their lifetime due to
subsequent recombination with majority electrons. Our results indicate
that boron is amphoteric in 4H-SiC, so that it could also affect p-type
material in undesirable ways. The donor transitions of $\textrm{B}_{\textrm{C}}$
and $V_{\textrm{C}}\textrm{B}_{\textrm{Si}}$ could render these defects
as trapping/scattering/compensating centers, with direct impact on
hole mobility, device frequency response and doping efficiency.

Formation energies of Al defects are shown in Fig.~\ref{fig1}(b).
Notably, and contrary to its boron counterpart, the formation energy
of $\textrm{Al}_{\textrm{C}}$ is very high. Although reference Al-C
and Al-Si bond energies are not very different (268~kJ/mol and 247~kJ/mol,
respectively \citep{Haynes2014}), they already hint for a preferential
formation of $\textrm{Al}_{\textrm{Si}}$-C bonds (against $\textrm{Al}_{\textrm{C}}$-Si).
The low stability of $\textrm{Al}_{\textrm{C}}$, however, can be
attributed to volumetric effects, that is, the Al atom (atomic radius
143~pm) fits well within the Si site (atomic radius 118~pm), while
it builds up a rather large compressive strain on the tiny C site
(atomic radius 77~pm).

When the Fermi energy is close to mid-gap, the formation energy of
$\textrm{Al}_{\textrm{C}}$ is nearly 6.5~eV higher than that of
$\textrm{Al}_{\textrm{Si}}$. This means that unlike $\textrm{B}_{\textrm{C}}$,
$\textrm{Al}_{\textrm{C}}$ impurities should be virtually inexistent
in as-grown epitaxial 4H-SiC. A similar conclusion had been previously
reached by Fukumoto for $\textrm{Al}_{\textrm{C}}$ in 3C-SiC \citep{Fukumoto1996}.

Based on the above results, we note that by ascribing the D-center
observed in boron-doped material to $\textrm{B}_{\textrm{C}}$, we
must conclude that the formation of an Al analogue is unlikely. As
far as we are aware, there has been no report regarding the observation
of a D-center-like hole trap in boron-free SiC by means of DLTS. In
that respect, our interpretation is consistent with the available
junction capacitance measurements.

On the other hand, such view seems to clash with optically-detected
magnetic resonance measurements of B-, Al- and Ga-doped 4H- and 6H-SiC
samples (see Ref.~\citep{DuijnArnold1998,Baranov1999}). Here, the
observation of Al- and Ga-related signals with gyromagnetic angular
dependence close to that of a signal attributed to the boron-related
D-center, led to the proposal of a three-member family of deep-acceptors
consisting of $V_{\textrm{C}}X_{\textrm{Si}}$ complexes, with $X$
being a group-III species (B, Al or Ga), all of which showing axial
symmetry along the main crystallographic axis \citep{DuijnArnold1998,Baranov1999}.

Judging from (i) the calculated transition levels of $V_{\textrm{C}}\textrm{B}_{\textrm{Si}}$
and $V_{\textrm{C}}\textrm{Al}_{\textrm{Si}}$ which, as shown in
Figs.~\ref{fig1}(a) and \ref{fig1}(b), are located in the upper
half of the gap and far from that of the D-center; (ii) from the calculated
$g$-tensor for the $V_{\textrm{C}}\textrm{B}_{\textrm{Si}}$, which
as reported in Ref.~\citep{Torres2022} differs markedly from the
paramagnetic measurements \citep{DuijnArnold1998,Baranov1999}; (iii)
from the range of Fermi energies for which the paramagnetic neutral
state of $V_{\textrm{C}}X_{\textrm{Si}}$ is stable ($E_{\textrm{F}}\gtrsim E_{\textrm{c}}-1\:\textrm{eV}$);
and (iv) from the small energy difference between all $V_{\textrm{C}}X_{\textrm{Si}}$
alignments in the crystal, which hardly explains the preferential
formation of an axial defect, we are led to the conclusion that the
\emph{deep} group-III paramagnetic centers reported in Refs.~\citep{Baranov1996b,DuijnArnold1998}
are not likely to be connected to any of the defects in this study.
Although the calculated gyromagnetic tensor and hyperfine coupling
constants of $\textrm{B}_{\textrm{C}}$ were recently found to account
well for the EPR measurements pertaining to the deep paramagnetic
boron center \citep{Torres2022}, as far as we are aware, apart from
their close transition energies, there is no evidence for a direct
connection between this paramagnetic/photoluminescent signal and the
DLTS hole trap referred to as D-center. This is an issue that calls
for further work and clarification.

\subsection{Capture coefficients of group-III acceptors\label{subsec:c-x}}

We now move on to the kinetics of carrier capture by boron- and aluminum-related
traps in 4H-SiC. We started by calculating the potential energy profiles
between initial and final electronic states ($E_{i}$ and $E_{f}$)
along the mode coordinate that joins $Q_{i}$ and $Q_{f}$. Figure~\ref{fig1}(c)
shows the results for the $\textrm{B}_{\textrm{Si}}^{-}+\textrm{h}^{+}\rightarrow\textrm{B}_{\textrm{Si}}^{0}$
transition, which involves hole capture at $\textrm{B}_{\textrm{Si}}^{-}$
on the $h$-site. Here we find $Q_{i}=\Delta Q=1.83\:\textrm{amu}^{1/2}\textrm{Å}$,
$Q_{f}=0$ and a trap depth of $\Delta E=E_{i}-E_{f}=0.32~\textrm{eV}$.
Other parameters, including mode frequencies ($\hbar\Omega_{i}$ and
$\hbar\Omega_{f}$) of each electronic state and capture barrier ($E_{\sigma}$)
were also evaluated at this point.

From the above results we could evaluate of the vibrational lineshape
function (Eq.~\ref{eq:lineshape}). Quantities deemed useful for
our discussion can be found in Table~\ref{tab2}. Figures~\ref{fig1}(c)-(f)
may assist the reader in their interpretation. Some defects, for instance
with very high formation energy (unlikely to form), or as we will
find, with a vanishing capture coefficients, are not discussed.

\begin{table*}
\caption{\label{tab2}Calculated quantities for the evaluation of capture coefficients
of selected hole and electron traps in this study. From left to right
we have: the reference defect state $X^{q}(Q_{0})$ specifying the
charge $q$ and reference ionic geometry of the defect $Q_{0}$ on
which $W_{if}$ was calculated, degeneracy of the transition $g$,
sub-lattice site of the defect, depth of the carrier trap $\Delta E$,
mode coordinate displacement $\Delta Q$, mode frequency of the final
electronic state $\Omega_{f}$ and respective Huang-Rhys factor $S_{f}$,
coupling factor $\varLambda$, and electron-phonon coupling matrix
element $W_{if}$.}

\begin{ruledtabular}
\begin{tabular}{ccccccccccc}
 & $X^{q}(Q_{0})$ & $g$ & Site & $\Delta E$ (eV) & $\Delta Q$ ($\textrm{amu}^{1/2}$Å) & $\hbar\Omega_{f}$ (meV) & $S_{f}$ & $\varLambda$ &  & $W_{if}$ ($\textrm{eV}/\textrm{amu}^{1/2}$Å)\tabularnewline
\hline 
\multirow{2}{*}{$\textrm{B}_{\textrm{Si}}^{-}+\textrm{h}^{+}\rightarrow\textrm{B}_{\textrm{Si}}^{0}$} & \multirow{2}{*}{$\textrm{B}_{\textrm{Si}}^{0}(Q_{f})$} & \multirow{2}{*}{4} & $k$ & 0.38 & 1.81 & 40.5 & 14 & 0.60 &  & $7.12\times10^{-2}$\tabularnewline
 &  &  & $h$ & 0.32 & 1.83 & 44.5 & 16 & 0.69 &  & $9.36\times10^{-2}$\tabularnewline
\multirow{2}{*}{$\textrm{B}_{\textrm{Si}}^{0}+\textrm{e}^{-}\rightarrow\textrm{B}_{\textrm{Si}}^{-}$} & \multirow{2}{*}{$\textrm{B}_{\textrm{Si}}^{0}(Q_{i})$} & \multirow{2}{*}{1} & $k$ & 2.82 & 1.81 & 78.0 & 28 & 0.43 &  & $4.35\times10^{-2}$\tabularnewline
 &  &  & $h$ & 2.86 & 1.83 & 75.7 & 27 & 0.42 &  & $5.65\times10^{-2}$\tabularnewline
\multirow{2}{*}{$\textrm{Al}_{\textrm{Si}}^{-}+\textrm{h}^{+}\rightarrow\textrm{Al}_{\textrm{Si}}^{0}$} & \multirow{2}{*}{$\textrm{Al}_{\textrm{Si}}^{0}(Q_{f})$} & \multirow{2}{*}{1} & $k$ & 0.19 & 0.27 & 53.3 & 0.4 & 0.10 &  & $1.00\times10^{-1}$\tabularnewline
 &  &  & $h$ & 0.23 & 0.25 & 53.3 & 0.3 & 0.08 &  & $9.42\times10^{-2}$\tabularnewline
\multirow{2}{*}{$\textrm{Al}_{\textrm{Si}}^{0}+\textrm{e}^{-}\rightarrow\textrm{Al}_{\textrm{Si}}^{-}$} & \multirow{2}{*}{$\textrm{Al}_{\textrm{Si}}^{0}(Q_{i})$} & \multirow{2}{*}{1} & $k$ & 3.01 & 0.27 & 62.4 & 0.5 & 0.01 &  & $1.08\times10^{-3}$\tabularnewline
 &  &  & $h$ & 2.97 & 0.25 & 62.9 & 0.4 & 0.01 &  & $9.25\times10^{-4}$\tabularnewline
\multirow{2}{*}{$\textrm{B}_{\textrm{C}}^{-}+\textrm{h}^{+}\rightarrow\textrm{B}_{\textrm{C}}^{0}$} & \multirow{2}{*}{$\textrm{B}_{\textrm{C}}^{0}(Q_{f})$} & \multirow{2}{*}{3} & $k$ & 0.69 & 0.80 & 59.7 & 4 & 0.26 &  & $7.29\times10^{-2}$\tabularnewline
 &  &  & $h$ & 0.74 & 0.83 & 58.3 & 4 & 0.25 &  & $6.79\times10^{-2}$\tabularnewline
\multirow{2}{*}{$\textrm{B}_{\textrm{C}}^{0}+\textrm{e}^{-}\rightarrow\textrm{B}_{\textrm{C}}^{-}$} & \multirow{2}{*}{$\textrm{B}_{\textrm{C}}^{0}(Q_{i})$} & \multirow{2}{*}{1} & $k$ & 2.51 & 0.80 & 68.6 & 5 & 0.11 &  & $9.91\times10^{-3}$\tabularnewline
 &  &  & $h$ & 2.46 & 0.83 & 65.9 & 4 & 0.11 &  & $4.41\times10^{-3}$\tabularnewline
\multirow{2}{*}{$\textrm{B}_{\textrm{C}}^{0}+\textrm{h}^{+}\rightarrow\textrm{B}_{\textrm{C}}^{+}$} & \multirow{2}{*}{$\textrm{B}_{\textrm{C}}^{+}(Q_{f})$} & \multirow{2}{*}{1} & $k$ & 0.20 & 0.75 & 54.4 & 3 & 0.47 &  & $5.89\times10^{-2}$\tabularnewline
 &  &  & $h$ & 0.22 & 0.78 & 53.2 & 3 & 0.46 &  & $4.80\times10^{-2}$\tabularnewline
\multirow{2}{*}{$\textrm{B}_{\textrm{C}}^{+}+\textrm{e}^{-}\rightarrow\textrm{B}_{\textrm{C}}^{0}$} & \multirow{2}{*}{$\textrm{B}_{\textrm{C}}^{+}(Q_{i})$} & \multirow{2}{*}{3} & $k$ & 3.01 & 0.75 & 65.0 & 4 & 0.08 &  & $7.46\times10^{-3}$\tabularnewline
 &  &  & $h$ & 2.98 & 0.78 & 62.8 & 4 & 0.08 &  & $4.80\times10^{-3}$\tabularnewline
\end{tabular}
\end{ruledtabular}

\end{table*}

\begin{figure}
\includegraphics[width=8.5cm]{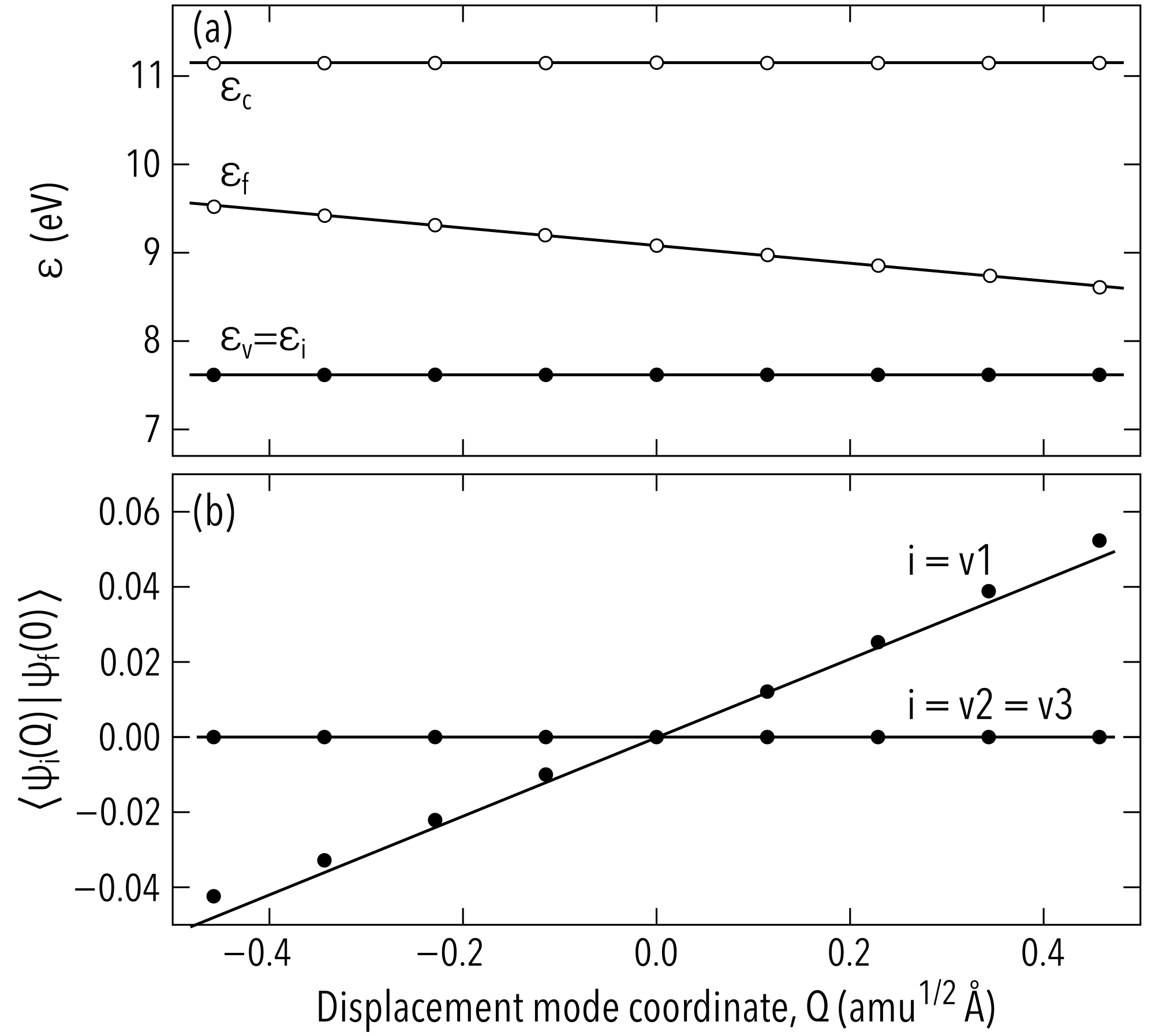}

\caption{\label{fig2}(a) Calculated Kohn-Sham eigenvalues of $\textrm{B}_{\textrm{Si}}^{0}(h)$
in 4H-SiC as a function of the displacement mode coordinate $Q$.
Only the minority spin channel energies are shown. Filled and empty
symbols represent occupied and unoccupied states, respectively. The
state with energy $\epsilon_{\textrm{f}}$ holds the trapped hole.
Eigenvalues of the topmost valence band and lowest conduction band
states are indicated as $\epsilon_{\textrm{v1}}$ and $\epsilon_{\textrm{c}}$,
respectively. (b) Overlap matrix elements $\left\langle \psi_{i}(Q)\left|\psi_{f}(Q=0)\right.\!\right\rangle $
with $i=\{\textrm{v}1,\,\textrm{v}2,\,\textrm{v}3\}$, for evaluating
the electron-phonon matrix element $W_{if}$ of $\textrm{B}_{\textrm{Si}}^{-}(h)+h^{+}\rightarrow\textrm{B}_{\textrm{Si}}^{0}(h)$
in 4H-SiC.}
\end{figure}

As prescribed in Refs.~\citep{Alkauskas2014,Turiansky2021}, electron-phonon
coupling matrix elements, $W_{if}$, should be evaluated at a reference
state where the single-particle defect level that accepts the hole
is well identified within the band gap. Given that $\textrm{B}_{\textrm{Si}}^{-}$
and $\textrm{Al}_{\textrm{Si}}^{-}$ states show a clean band gap,
neutral states were picked up as references for studying acceptor
transitions. Figure~\ref{fig2}(a) shows the change of the lowest
unoccupied one-electron state of $\textrm{B}_{\textrm{Si}}^{0}$ with
energy $\epsilon_{f}$ as a function of the mode coordinate $Q$.
Valence band top and conduction band bottom Kohn-Sham states ($\epsilon_{\textrm{v}}$
and $\epsilon_{\textrm{c}}$) are also represented in the same figure
as closed and open symbols, respectively. The hole state induced by
the defect is clearly within the gap and its energy change is approximately
linear with $Q$. The positive charge state of $\textrm{B}_{\textrm{C}}$
also has a deep and well localized one-electron level in the band
gap, and for that reason it was chosen for studying its donor transition.

The $\textrm{B}_{\textrm{Si}}^{-}+\textrm{h}^{+}\rightarrow\textrm{B}_{\textrm{Si}}^{0}$
transition involves the capture of a hole by a four-fold coordinated
$\textrm{B}_{\textrm{Si}}^{-}$ impurity, which must be concomitant
with the displacement of the B atom away from one of its four carbon
neighbors. Considering an approximate local tetrahedral symmetry for
$\textrm{B}_{\textrm{Si}}^{-}$, there are four possible routes for
this to happen, and hence $g=4$ for this transition (see third column
of Tab.~\ref{tab2}). For the case of $\textrm{B}_{\textrm{C}}^{-}+\textrm{h}^{+}\rightarrow\textrm{B}_{\textrm{C}}^{0}$
the final state is one of three possible Jahn-Teller split configurations
($g=3$), and the same applies to the $\textrm{B}_{\textrm{C}}^{+}+\textrm{e}^{-}\rightarrow\textrm{B}_{\textrm{C}}^{0}$
transition. All other carrier capture reactions have a unique final
state ($g=1$).

The depths of carrier traps ($\Delta E$) listed in the fifth column
of Tab.~\ref{tab2}, were already discussed in the previous section.
Here we underline that when we are dealing with hole or electron traps,
this quantity refers to a transition level $\Delta E$ above or below
the valence or conduction band edges, respectively. Regarding the
mode coordinate displacements ($\Delta Q$) we find important differences
between $\textrm{B}_{\textrm{Si}}$ ($\Delta Q\sim1.8$~$\textrm{amu}^{1/2}$Å)
and other defects, most notably $\textrm{Al}_{\textrm{Si}}$ ($\Delta Q<0.3$~$\textrm{amu}^{1/2}$Å)
shallower acceptor. This is a consequence of the large displacement
of the B atom and one of its C neighbors after capturing a hole at
a localized state ($\textrm{B}_{\textrm{Si}}^{-}+\textrm{h}^{+}\rightarrow\textrm{B}_{\textrm{Si}}^{0}$).
In contrast, the minute displacement of Al and its immediate neighbors
upon hole capture into a relatively diffuse state, hints for an effective-mass
theory shallow acceptor. This limits the scope for the analysis of
$\textrm{Al}_{\textrm{Si}}^{-}+\textrm{h}^{+}\rightarrow\textrm{Al}_{\textrm{Si}}^{0}$
transitions within the present framework.

The above difference in $\Delta Q$ values is illustrated in Figs.~\ref{fig1}(c)
and \ref{fig1}(d). The effect is also manifested in electron capture
reactions involving initial states represented by the upper parabolas
in the same figures, and has critical consequences to the underlying
transition mechanism governing the capture of carriers by $\textrm{B}_{\textrm{Si}}$
and $\textrm{Al}_{\textrm{Si}}$ impurities. The Huang-Rhys factor,

\begin{equation}
S_{f}=\frac{\Delta Q^{2}\Omega_{f}}{2\hbar},
\end{equation}
was calculated for hole capture by both $\textrm{B}_{\textrm{Si}}^{-}$
and $\textrm{Al}_{\textrm{Si}}^{-}$ using the final ground state
frequency $\Omega_{f}$ (see seventh column of Tab.~\ref{tab2}).
This choice is justified by the fact that we are addressing a capture
process (not emission). The harmonic mode frequencies for the initial
states ($\Omega_{i}$) can also be obtained by inspection of $\Omega_{f}$
values in Tab.~\ref{tab2}. For instance, $\Omega_{i}$ for reaction
$\textrm{B}_{\textrm{Si}}^{-}+\textrm{h}^{+}\rightarrow\textrm{B}_{\textrm{Si}}^{0}$
is equivalent to $\Omega_{f}$ for reaction $\textrm{B}_{\textrm{Si}}^{0}+\textrm{e}^{-}\rightarrow\textrm{B}_{\textrm{Si}}^{-}$.
Notably, $\Omega_{i}$ frequencies are invariably higher than their
final state counterparts ($\Omega_{f}$) in hole capture processes.
We interpret this result as reflecting a softening of impurity-ligand
bonds upon hole capture. The opposite (bond strengthening) occurs
upon electron capture, suggesting that the carriers occupy electronic
states with bonding character.

The calculated $S_{f}$ values are also reported in Tab.~\ref{tab2}.
Transitions involving $\textrm{B}_{\textrm{Si}}$ show $S_{f}\gg1$
which is characteristic of a deep trap with a rather strong electron-phonon
coupling. Conversely, hole capture by $\textrm{Al}_{\textrm{Si}}^{-}$
shows $S_{f}<1$, indicating weak electron-phonon coupling, typical
of effective-mass-like impurities, again anticipating that for this
case, Eq.~\ref{eq:c} --- which assumes that the energy is dissipated
via multi-phonon emission --- is not applicable.

A clearer classification of the coupling strength within the harmonic
approximation is obtained with help of a \emph{coupling factor} as
defined by Stoneham \citep{Stoneham1981}, $\varLambda=S_{f}/(S_{f}+p)$
with $p=\Delta E/\hbar\Omega_{f}$ (see ninth column of \ref{tab2}).
For $\varLambda=0$, \emph{i.e.} no coupling, the potential energy
surfaces of initial and final states are nested; When $\varLambda=1/2$
the coupling is strong --- the potential energy curve of the final
state crosses initial state at its energy minimum and there is no
capture barrier; When $\varLambda\gg1/2$ the coupling is very strong,
and the crossing between the potential energy curves occurs between
the energy minima of initial and final states. The later case is typical
of bistable defects and there can be substantial energy barriers separating
the initial from the final states.

As shown by Fig~\ref{fig1}(d), there is no crossing between $\textrm{Al}_{\textrm{Si}}^{-}+\textrm{h}^{+}$
and $\textrm{Al}_{\textrm{Si}}^{0}$ potential curves (nested potential
energy curves) and according to Tab.~\ref{tab2}, the coupling factor
for transitions involving $\textrm{Al}_{\textrm{Si}}$ indicates a
weak coupling ($\varLambda\lesssim0.1$). On the other hand, transitions
involving $\textrm{B}_{\textrm{Si}}$ show a strong coupling factor
close to $\varLambda\sim0.5,$ anticipating a minute capture barrier.
We must take the calculated barriers, especially the shallower ones
($E_{\sigma}\lesssim0.1\:\textrm{eV}$) with caution. Variations of
$\sim0.05~\textrm{eV}$ in the barrier can change the calculated capture
rates by one order of magnitude. Such variations could be easily induced
by approximations, such as neglecting changes in entropy across the
barrier or the adoption of a single effective mode to couple the initial
and final states.

Table.~\ref{tab2} also reports intermediate coupling transitions
for $\textrm{B}_{\textrm{C}}^{-}+\textrm{h}^{+}\rightarrow\textrm{B}_{\textrm{C}}^{0}$
and strong coupling for $\textrm{B}_{\textrm{C}}^{0}+\textrm{h}^{+}\rightarrow\textrm{B}_{\textrm{C}}^{+}$.
Other transitions show a weak coupling character, such as for the
case of electron capture at $\textrm{B}_{\textrm{C}}^{0}+\textrm{e}^{-}\rightarrow\textrm{B}_{\textrm{C}}^{-}$
and $\textrm{B}_{\textrm{C}}^{+}+\textrm{e}^{-}\rightarrow\textrm{B}_{\textrm{C}}^{0}$,
suggesting high capture barriers and therefore slow capture kinetics,
at least in the form of non-radiative transitions via multi-phonon
emission.

In Fig.~\ref{fig2}(b) we plot the overlap integrals $\langle\psi_{i}(Q)|\psi_{f}(0)\rangle$
for the hole capture reaction $\textrm{B}_{\textrm{Si}}^{-}(h)+h^{+}\rightarrow\textrm{B}_{\textrm{Si}}^{0}(h)$.
Identical calculations were carried out for all transitions listed
in Tab.~\ref{tab2}. Like in Ref.~\citep{Alkauskas2014}, we found
three topmost valence band states and only one conduction band minimum
at $\mathbf{k}=\Gamma$. Three possible transitions were therefore
considered for hole capture, differing only on the choice of the initial
one-electron state $\psi_{i}$ describing the free hole. These upper
valence band states are labelled as $i=\{\textrm{v}1,\,\textrm{v}2,\,\textrm{v}3\}$.
The final state is $\textrm{B}_{\textrm{Si}}^{0}(h)$ at $Q=0$, which
is the ground state after hole capture, and importantly, shows a well
defined hole-trapping level within the band gap. The derivative of
the overlap integral was used to find the transition matrix element,

\begin{equation}
W_{if}=(\epsilon_{f}-\epsilon_{i})\Bigl\langle\frac{\partial\psi_{i}}{\partial Q}\Bigl|\psi_{f}(0)\Bigr\rangle,
\end{equation}
which is evaluated from the slope of $\langle\psi_{i}(Q)|\psi_{f}(0)\rangle$
in Fig.~\ref{fig2}(b) around $Q=0$. The coupling to the $\psi_{\textrm{v1}}$
state is several orders of magnitude higher than for $\psi_{\textrm{v2}}$
and $\psi_{\textrm{v3}}$. The effective value of $W_{if}$ is defined
as the mean-square average from the three individual matrix elements,
and it is dominated by the term with $i=\textrm{v1}$. The resulting
transition matrix elements are listed in the last column of Tab.~\ref{tab2}.

\begin{figure}
\includegraphics[width=8.5cm]{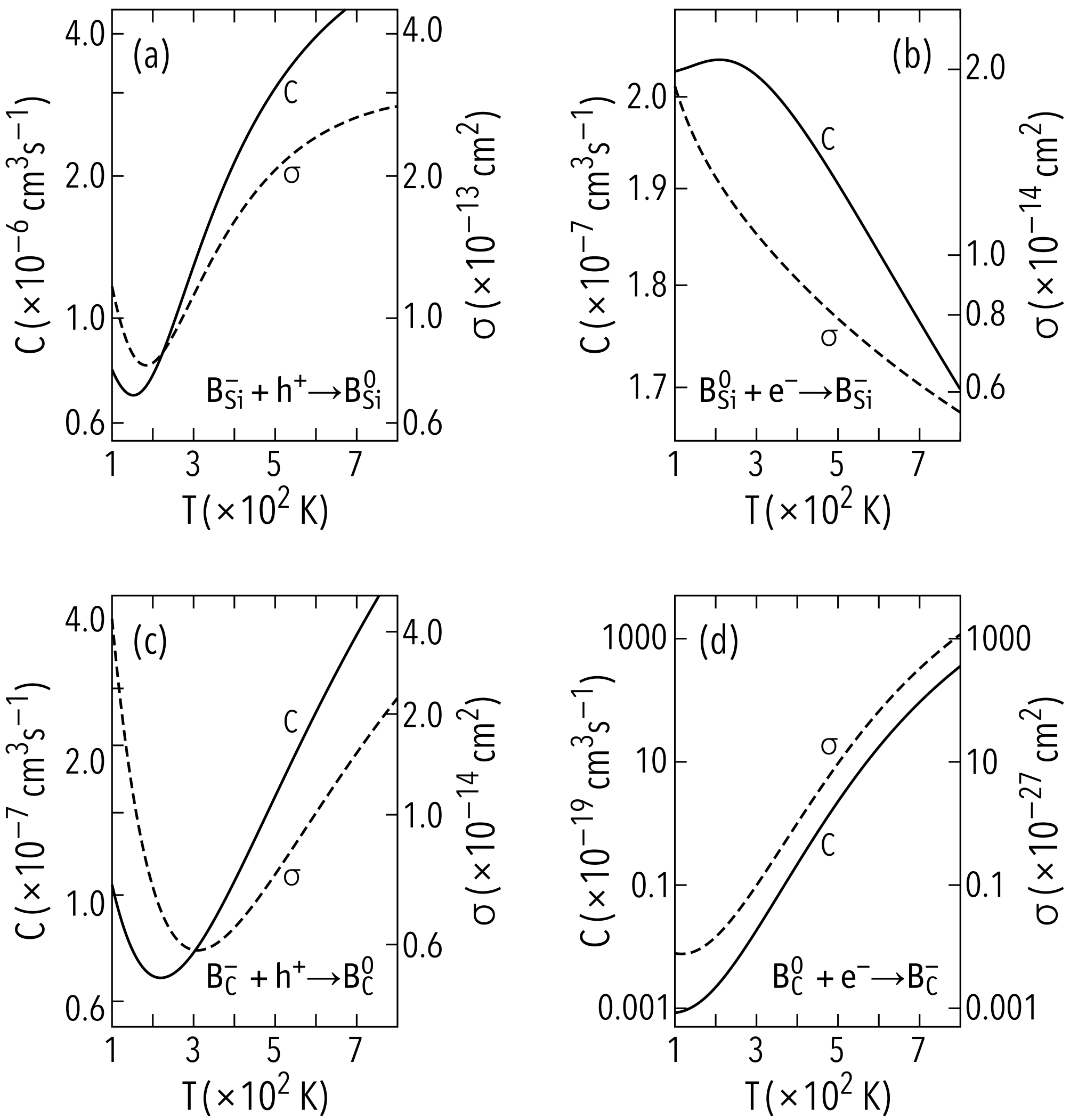}

\caption{\label{fig3}Calculated temperature dependence of the carrier capture
coefficients ($C$) and cross sections ($\sigma$) for (a) $\textrm{B}_{\textrm{Si}}^{-}+\textrm{h}^{+}\rightarrow\textrm{B}_{\textrm{Si}}^{0}$,
(b) $\textrm{B}_{\textrm{Si}}^{0}+\textrm{e}^{-}\rightarrow\textrm{B}_{\textrm{Si}}^{-}$,
(c) $\textrm{B}_{\textrm{C}}^{-}+\textrm{h}^{+}\rightarrow\textrm{B}_{\textrm{C}}^{0}$,
and (d) $\textrm{B}_{\textrm{C}}^{0}+\textrm{e}^{-}\rightarrow\textrm{B}_{\textrm{C}}^{-}$
in 4H-SiC.}
\end{figure}

Figure~\ref{fig3} shows the calculated capture coefficients as solid
lines referred to the left-vertical axes of the graphs. These were
obtained from Eq.~\ref{eq:c} in a range of temperatures from 100~K
up to 700~K, typically used in DLTS scans.

As indicated in Refs.~\citep{Henry1977,Stoneham1981}, the rate of
non-radiative capture follows a power law with the respect to the
energy drop, leading to large capture coefficients of up to $\sim10^{-6}\:\textrm{cm}^{3}\textrm{s}^{-1}$,
which in the case of 4H-SiC, corresponds to capture cross-sections
of about $10^{-14}\textrm{-}10^{-13}\:\textrm{cm}^{2}$. Radiative
transitions are usually several orders of magnitude slower.

Table~\ref{tab3} reports the calculated cross-sections for the capture
of electrons ($\sigma_{n}$) and holes ($\sigma_{p}$) by boron and
aluminum defects in 4H-SiC at different sub-lattice sites ($k$ and
$h$). Values are given for several temperatures in the range 100-700~K.
Figure~\ref{fig3} shows the temperature dependence of the transition
rate for four representative cases, respectively for hole and electron
capture by $\textrm{B}_{\textrm{Si}}^{-}$ and $\textrm{B}_{\textrm{Si}}^{0}$
{[}insets (a) and (b){]}, and hole and electron capture by $\textrm{B}_{\textrm{C}}^{-}$
and $\textrm{B}_{\textrm{C}}^{0}$ {[}insets (c) and (d){]}.

Looking at Fig.~\ref{fig3}, we should bear in mind two fundamental
processes driving a transition from the initial state (empty trap
plus free carrier) to the final state (bound carrier). The first is
that with raising the temperature, higher energy vibrational states
in the initial state become populated, increasing the probability
of overcoming the capture barrier $E_{\sigma}$, thus raising the
transition rate. This effect is most important on transitions involving
capture barriers of up to few tenths of eV like for $\textrm{B}_{\textrm{Si}}^{-}+\textrm{h}^{+}\rightarrow\textrm{B}_{\textrm{Si}}^{0}$
and $\textrm{B}_{\textrm{C}}^{0}+\textrm{e}^{-}\rightarrow\textrm{B}_{\textrm{C}}^{-}$
in Figs.~\ref{fig3}(a) and (d), respectively.

The second process relates to scattering --- with raising the temperature
and consequently the average thermal velocity of carriers, the capture
probability decreases. This is most visible for the case of capture
by neutral traps with small capture barriers (like $\textrm{B}_{\textrm{Si}}^{0}+\textrm{e}^{-}\rightarrow\textrm{B}_{\textrm{Si}}^{-}$).
However, when traps are attractive, \emph{i.e.} their charge state
is opposite to the charge of the free carrier, we have to consider
a capture enhancement due to long-range Coulomb attraction. This is
not possible to account from first-principles within the limits of
the supercell size. Instead, we employ the Sommerfeld scaling factor
$f_{\textrm{s}}(T)$ in Eq.~\ref{eq:c} as described in Ref.~\citep{Turiansky2021}.
The effect is visible in the low temperature regions of Figs.~\ref{fig3}(a)
and (c) --- in the high temperature regime, we observe an Arrhenius-like
behavior typical of thermal activation involving the capture barrier
(first process).

\begin{table*}
\caption{\label{tab3}Calculated cross sections for capture of electrons and
holes by boron and aluminum defects in 4H-SiC at different sub-lattice
sites ($k$ and $h$) and selected temperatures. All cross sections
($\sigma$) are given in $\textrm{cm}^{2}$. The right-most column
indicates the order of magnitude (OM) of $\sigma$ values.}

\begin{ruledtabular}
\begin{tabular}{lllccccc}
 &  &  & \multicolumn{4}{c}{Capture cross section, $\sigma(T)$} & \tabularnewline
\cline{4-7}
Reaction & site & $E_{\sigma}$ & $T=100\:\textrm{K}$ & $T=300\:\textrm{K}$ & $T=500\:\textrm{K}$ & $T=700\:\textrm{K}$ & OM\tabularnewline
\hline 
\multirow{2}{*}{$\textrm{B}_{\textrm{Si}}^{-}+\textrm{h}^{+}\rightarrow\textrm{B}_{\textrm{Si}}^{0}$} & $k$ & 0.06 & 14 & 10 & 11 & 12 & \multirow{2}{*}{$\times10^{-13}$}\tabularnewline
 & $h$ & 0.13 & 1.2 & 1.1 & 2.1 & 2.7 & \tabularnewline
\multirow{2}{*}{$\textrm{B}_{\textrm{Si}}^{0}+\textrm{e}^{-}\rightarrow\textrm{B}_{\textrm{Si}}^{-}$} & $k$ & 0.00 & 1.2 & 0.7 & 0.4 & 0.3 & \multirow{2}{*}{$\times10^{-14}$}\tabularnewline
 & $h$ & 0.01 & 1.8 & 1.1 & 0.8 & 0.6 & \tabularnewline
\multirow{2}{*}{$\textrm{B}_{\textrm{C}}^{-}+\textrm{h}^{+}\rightarrow\textrm{B}_{\textrm{C}}^{0}$} & $k$ & 0.29 & 13 & 5.9 & 6.3 & 7.5 & \multirow{2}{*}{$\times10^{-14}$}\tabularnewline
 & $h$ & 0.32 & 6.1 & 3.0 & 3.5 & 4.6 & \tabularnewline
\multirow{2}{*}{$\textrm{B}_{\textrm{C}}^{0}+\textrm{e}^{-}\rightarrow\textrm{B}_{\textrm{C}}^{-}$} & $k$ & 1.48 & 0.0 & 1.2 & 98 & 3200 & \multirow{2}{*}{$\times10^{-27}$}\tabularnewline
 & $h$ & 1.54 & 0.0 & 0.1 & 9.5 & 327 & \tabularnewline
\multirow{2}{*}{$\textrm{B}_{\textrm{C}}^{0}+\textrm{h}^{+}\rightarrow\textrm{B}_{\textrm{C}}^{+}$} & $k$ & 0.00 & 15 & 8.5 & 5.8 & 4.3 & \multirow{2}{*}{$\times10^{-14}$}\tabularnewline
 & $h$ & 0.00 & 10 & 5.9 & 4.1 & 3.1 & \tabularnewline
\multirow{2}{*}{$\textrm{B}_{\textrm{C}}^{+}+\textrm{e}^{-}\rightarrow\textrm{B}_{\textrm{C}}^{0}$} & $k$ & 2.08 & 0.0 & 0.0 & 20 & 1200 & \multirow{2}{*}{$\times10^{-30}$}\tabularnewline
 & $h$ & 2.13 & 0.0 & 0.0 & 8.4 & 460 & \tabularnewline
\end{tabular}
\end{ruledtabular}

\end{table*}

Regarding the data presented in Table~\ref{tab3}, we do not include
the capture rates of either $\textrm{Al}_{\textrm{Si}}$- or $\textrm{Al}_{\textrm{C}}$-related
transitions. In the first case the center has the characteristics
of an effective-mass center and should not be analyzed within the
present method. In the second case the formation energy of $\textrm{Al}_{\textrm{C}}$
(compared to $\textrm{Al}_{\textrm{Si}}$) was considered sufficiently
high to render it irrelevant.

Importantly, the results suggest that $\textrm{B}_{\textrm{Si}}^{-}$
can act as a non-radiative recombination center in n-type 4H-SiC,
showing relatively large capture cross-sections ($\sim10^{-14}\textrm{-}10^{-13}\:\textrm{cm}^{2}$)
for both $\textrm{B}_{\textrm{Si}}^{-}+\textrm{h}^{+}\rightarrow\textrm{B}_{\textrm{Si}}^{0}$
and $\textrm{B}_{\textrm{Si}}^{0}+\textrm{e}^{-}\rightarrow\textrm{B}_{\textrm{Si}}^{-}$
transitions. We may understand the recombination mechanism as analogous
to the ``bistability-mediated carrier recombination'' proposed by
Du and co-workers for the $\textrm{BO}_{2}$ complex in Si (responsible
for light-induced degradation of Si solar cells) \citep{Du2006}.
It involves a two-step process in which, immediately after hole capture
by $\textrm{B}_{\textrm{Si}}^{-}$, the center transforms into a configuration
featuring a dangling bond whose wave function extends into an interstitial
region of the crystal and therefore overlaps with conduction-band
states, thereby facilitating electron capture.

Carbon substitutions, respectively $\textrm{B}_{\textrm{C}}^{-}$
and $\textrm{B}_{\textrm{C}}^{0}$, are effective hole traps with
cross-sections of $\sim10^{-14}\:\textrm{cm}^{2}$. We do not find
strong interactions between these defect states and free electrons.
Their large capture barriers lead to vanishing capture coefficients.
This finding is at odds with the hypothesis that $\textrm{B}_{\textrm{C}}$
shows strong recombination activity in n-type 4H-SiC. Instead, it
suggests that $\textrm{B}_{\textrm{C}}$ is a trapping center without
significant impact on the minority carrier life-time. Of course, this
does not preclude recombination from occurring via alternative pathways
that were not considered here, such as trap-assisted Auger recombination.
For instance, the capture of two holes by $\textrm{B}_{\textrm{C}}$
with formation of $\textrm{B}_{\textrm{C}}^{+}(e^{\uparrow\uparrow++})$
could attract minority electrons into an excited state induced by
the defect, $\textrm{B}_{\textrm{C}}^{0}(e^{\uparrow\uparrow++}+\textrm{e}^{-*})$,
and subsequent hole emission assisted by recombination. Trap-assisted
Auger recombination rates scale as the second power of the minority
carrier density. It is negligible compared to multi-phonon relaxation
under low injection/excitation levels, but could be the main limiting
factor in the presence of a high density of steady carrier density
($\sim10^{18}\:\textrm{cm}^{-3}$) \citep{Zhao2023}. Such processes
involve complex many-body interactions and are beyond the scope of
the present work.

On the other hand, the calculation of a slow electron capture rate
by $\textrm{B}_{\textrm{C}}$ defects is consistent with reports of
radiative recombination related to boron, as reported after photoluminescence
measurements \citep{Kuwabara1975,Sridhara1998}. Clarifying the role
of alternative recombination pathways therefore constitutes an important
direction for future work.

\section{Conclusions\label{sec:conclusions}}

We reviewed the current state-of-the-art regarding the modeling of
the electronic structure of p-type dopants in 4H-SiC. Previous results
were revisited in light of modern computational methodologies based
on hybrid density functional theory, which overcome the limitations
associated with the underestimation of the band gap in conventional
density functional approaches. The calculations include defect formation
energies and electronic transition levels. We further investigated
the kinetics of carrier trapping by p-type dopant-related defects
within the framework of the multi-phonon emission mechanism.

We show that the most probable boron and aluminum impurities in 4H-SiC
are strikingly different, not only in terms of local geometries and
site locations, but also regarding their interactions with minority
and majority carriers. While $\textrm{B}_{\textrm{Si}}$ and $\textrm{B}_{\textrm{C}}$
have very similar formation energies, consistent with the observation
of both defects in as-grown SiC, we confirm previous indications that
$\textrm{Al}_{\textrm{C}}$ is unlikely to be found in 4H-SiC. We
find that under intrinsic conditions the formation energy of $\textrm{Al}_{\textrm{C}}$
is approximately 6.5 eV higher than that of $\textrm{Al}_{\textrm{Si}}$,
effectively precluding the formation of the former.

An important finding is that $\textrm{B}_{\textrm{Si}}$ shows strong
electron-phonon coupling, while $\textrm{Al}_{\textrm{Si}}$ shows
weak electron-phonon coupling typical of effective-mass shallow acceptors.
This renders the former highly efficient at capturing both electrons
and holes, making it a plausible contributor to minority-carrier lifetime
degradation of n-type 4H-SiC, where the presence of boron could be
unintentional (contamination during material growth or device processing).

We also predict the existence of a donor transition for $\textrm{B}_{\textrm{C}}$.
If confirmed experimentally, this finding could present an additional
challenge for p-type doping. Besides being ineffective as a dopant,
$\textrm{B}_{\textrm{C}}$ could trap up to two free holes, with potentially
detrimental consequences for device performance in terms of carrier
transport and dopant activation of p-type layers.
\begin{acknowledgments}
Support by the FCT is acknowledged through projects UID/50025/2025
and LA/P/0037/2020. The author acknowledges the High Performance Computing
Chair of the University of Évora for providing computing time in the
OBLIVION supercomputer under project reference 2024.HPCC.008.
\end{acknowledgments}

\section*{Data Availability}

The data that support the findings of this study are available from
the corresponding author upon request.\bibliographystyle{apsrev4-2}
\bibliography{refs}

\end{document}